\documentclass[twocolumn,english,prl]{revtex4}
\usepackage[T1]{fontenc}
\usepackage[latin9]{inputenc}
\usepackage[active]{srcltx}
\usepackage{bm}
\usepackage{amsmath}
\usepackage{amssymb}
\usepackage{stackrel}
\usepackage{graphicx}

\makeatletter
\@ifundefined{textcolor}{}
{%
 \definecolor{BLACK}{gray}{0}
 \definecolor{WHITE}{gray}{1}
 \definecolor{RED}{rgb}{1,0,0}
 \definecolor{GREEN}{rgb}{0,1,0}
 \definecolor{BLUE}{rgb}{0,0,1}
 \definecolor{CYAN}{cmyk}{1,0,0,0}
 \definecolor{MAGENTA}{cmyk}{0,1,0,0}
 \definecolor{YELLOW}{cmyk}{0,0,1,0}
}

\usepackage[colorlinks,linkcolor=blue,anchorcolor=blue,citecolor=blue]{hyperref}

\makeatother

\usepackage{babel}
\begin{document}

\title{Nonequilibrium Steady States and Resonant Tunneling in Time-Periodically
Driven Systems with Interactions}

\author{Tao Qin}

\affiliation{Institut für Theoretische Physik, Goethe-Universität, 60438 Frankfurt/Main,
Germany}

\author{Walter Hofstetter}

\affiliation{Institut für Theoretische Physik, Goethe-Universität, 60438 Frankfurt/Main,
Germany}
\begin{abstract}
Time-periodically driven systems are a versatile toolbox for realizing
interesting effective Hamiltonians. Heating, caused by excitations
to high-energy states, is a challenge for experiments. While most
setups address the relatively weakly-interacting regime so far, it
is of general interest to study heating in strongly correlated systems.
Using Floquet dynamical mean-field theory, we study non-equilibrium
steady states (NESS) in the Falicov-Kimball model, with time-periodically
driven kinetic energy or interaction. We systematically investigate
the nonequilibrium properties of the NESS. For a driven kinetic energy,
we show that resonant tunneling, where the interaction is an integer
multiple of the driving frequency, plays an important role in the
heating. In the strongly correlated regime, we show that this can
be well understood using Fermi\textquoteright s golden rule and the
Schrieffer-Wolff transformation for a time-periodically driven system.
We furthermore demonstrate that resonant tunneling can be used to
control the population of Floquet states to achieve ``photo-doping''. 
For driven interactions introduced by an oscillating magnetic field
near a Feshbach resonance widely adopted, we find that the double occupancy is strongly
modulated. Our calculations apply to shaken ultracold atom systems, and to
solid state systems in a spatially uniform but time-dependent electric
field. They are also closely related to lattice modulation spectroscopy. Our calculations are helpful to understand the latest experiments
on strongly correlated Floquet systems. 
\end{abstract}
\maketitle

\section{Introduction}

Time-periodically driven ultra-cold quantum gases are a highly promising
platform for realizing topologically non-trivial Hamiltonians. Two
paradigmatic models, the Harper-Hofstadter and Haldane Hamiltonians,
have been realized experimentally~\cite{Bloch2013,Ketterle2013,Jotzu:2014fk}.
Meanwhile, heating, where atoms are excited to higher energy states,
is unavoidable. This issue is even more relevant~\cite{Jotzu2015a}
for a time-periodically driven, strongly interacting system. A driven
system can be closed or open. For a closed driven system, generally
it can be shown that an isolated, driven many-body system will be
heated up to infinite temperature in the long-time limit~\cite{DAlessio2014,Kuwahara2016fmt}.
It has been analytically proven that the linear response energy absorption
rate decays exponentially as a function of driving frequency for a
system with local interactions~\cite{Abanin2015esh}, and that the
effective Hamiltonian~\cite{bukov2015} can describe the transient
dynamics on a certain time scale~\cite{Kuwahara2016fmt}. An open
driven system, when coupled to a bath, can reach a nonequilibrium
steady state (NESS)~\cite{Aoki2009,Vorberg2013gbe,Seetharam2015cb},
which is fundamentally different from an equilibrium phase. Here we
take this latter path to study NESS of a many-body system subjected
to different kinds of driving. 

Consider an ultracold atomic system in an optical lattice, $H=H_{K}+H_{U}$,
where the Hamiltonian $H$ consists of the kinetic part $H_{K}$ and
the interacting part $H_{U}$. With time-periodic driving, the interplay
between interactions and driving frequency may lead to resonant tunneling,
which can be detrimental to experiments, because many atoms are excited
away from the ground state. On the other hand, it can also be useful
if it is well-controllable, because it can realize ``photo-doping''~\cite{Iwai2003uos},
an analog to real doping in solid state systems. Different ways have
been designed to drive the system in order to achieve different goals~\cite{Eckardt2017aqg}.
Raman-laser assisted tunneling~\cite{Jaksch2003,Bloch2013,Ketterle2013}
and lattice shaking~\cite{Struck19082011,Jotzu:2014fk,Flaschner:2016ab,Jotzu2015a},
which are a common strategy for realizing systems with artificial
gauge fields, can be regarded as driving of the kinetic energy $H_{K}$.
In Ref.~\cite{Desbuquois2017ctf}, by creating an array of double
wells occupied by pairs of interacting fermionic atoms, it has been
demonstrated that resonant tunneling, where the interaction is equal
to the driving frequency, leads to higher double occupancy. In Ref.~\cite{Gorg:2018aa},
a time periodically driven hexagonal lattice is studied. This experiment
study shows that a driven, strongly correlated fermionic system can
be described by an effective Hamiltonian in the high frequency regime,
and observes a resonance of double occupancy and a sign reversal of
magnetic tunneling with increasing interaction $U$. On the other
hand, an oscillating magnetic field near a Feshbach resonance is widely
used as a Feshbach resonance management to investigate Bose-Einstein
condensates~\cite{Kevrekidis2003frm,Abdullaev2003abe,Gong2009mbc},
and the Fermi-Hubbard model~\cite{Dirks2015fms}, and has been adopted experimentally~\cite{Meinert2016fec}
to realize density dependent tunneling~\cite{Rapp2012ulg,Meinert2016fec},
which is essentially a driving of the interaction. These two driving
protocols are equivalent to some degree, but they are generally different
in many aspects. Moreover, systems subjected to double modulation~\cite{Greschner2014euh,Greschner2014dds},
i.e. driving $H_{K}$ and $H_{U}$ at the same time, are under intense
investigation. 

However, most theoretical studies are based on effective Hamiltonians
and low-dimensional systems, where the introduction of driving only
leads to an extra parameter within the effective Hamiltonian, and
essentially an equilibrium state is considered. Because a time-periodically
driven system is generally out-of-equilibrium, a more realistic description
should go beyond the equilibrium state. On the other hand, in the
context of nonequilibrium physics, the fermionic Falicov-Kimball model,
a limiting case of the Hubbard model, is under intense study~\cite{Eckstein2008nss,Kollar2011gbe,Aoki2008,Aoki2009,Aoki2014,Herrmann2016ndc}.
It provides a good testing ground for our understanding of nonequilibrium
physics. 

In this manuscript, we use the recently developed Floquet dynamical
mean field theory (DMFT)~\cite{Freericks2006,Aoki2008,Freericks2008,Freericks2008nds,Tsuji2014}
and its generalization to a driven interaction, to systematically
study properties of NESS of a fermionic
Falicov-Kimball model, with driven kinetic energy or driven interaction.
We clearly demonstrate nonequilibrium properties of the NESS. For
a driven kinetic energy, we show that resonant tunneling can lead
to heating, and that it can be used to implement ``photo-doping''.
In the strongly correlated regime, we find that our results can be
well explained using Fermi's golden rule~\cite{Shirley1965sse} and
the Schrieffer-Wolff transformation~\cite{Schrieffer1966rba,Bukov2016sf}
of a time-periodically driven system. Our prediction of a resonance
of the double occupancy has been observed in a recent experiment~\cite{Gorg:2017aa}.
For a driven interaction, we find very interesting behavior of the
double occupancy as a function of driving amplitude. Our calculations are also useful for understanding lattice
modulation spectroscopy~\cite{Stoferle2004ti,Jordens:2008aa,Greif2011pl},
first introduced in Ref.~\cite{Stoferle2004ti}, and theoretically
investigated in Ref.~\cite{Dirks2014tdc}, in which excitations are
created by periodically modulating the intensity of one or more laser
beams creating the optical lattice potential. The modulation introduces
a time-periodic perturbation of both the kinetic and interaction energy. 

This manuscript is organized as follows. In Sec.~\ref{sec:Floquet-DMFT},
we give a short introduction to the Floquet DMFT formalism. In Sec.~\ref{sec:Physical-quantities},
we present details of the physical quantities we plan to study. In
Sec.~\ref{sec:Results-and-discussions}, we present results of our
numerical simulations and discuss properties of the NESS in a driven
Falicov-Kimball model. Sec.~\ref{sec:Conclusions} presents the conclusion
and future directions. 

\section{\label{sec:Floquet-DMFT}Floquet DMFT formalism}

We give a brief review of Floquet's theorem. For a time-periodically
driven system, the Hamiltonian is time-periodic $H\left(t\right)=H\left(t+\mathcal{T}\right)$
where $\mathcal{T}=\frac{2\pi}{\Omega}$ with $\Omega$ the driving
frequency. According to Floquet's theorem~\cite{Aoki2014}, the solution
$\psi_{\alpha}\left(t\right)$, where $\alpha$ is a quantum number,
to the Schr\"odinger equation has the form $\psi_{\alpha}\left(t\right)=e^{-\mathrm{i}\epsilon_{\alpha}t}u_{\alpha}\left(t\right)$
with $u_{\alpha}\left(t\right)=u_{\alpha}\left(t+\mathcal{T}\right)$.
$u_{\alpha}\left(t\right)$ can be Fourier expanded as $u_{\alpha}\left(t\right)=\sum_{n=-\infty}^{\infty}u_{\alpha}^{n}e^{-\mathrm{i}n\Omega t}$.
From the Schr\"odinger equation, one obtains $\sum_{n}H_{mn}u_{\alpha}^{n}=\left(\epsilon_{\alpha}+m\Omega\right)u_{\alpha}^{m}$
with $H_{mn}=\frac{1}{\mathcal{T}}\int_{0}^{\mathcal{T}}dte^{\mathrm{i}\left(m-n\right)\Omega t}H\left(t\right)$.
$\epsilon_{\alpha}+m\Omega$ is denoted as quasi-energy and is not
bounded. 

Floquet DMFT is a non-perturbative method for studying the NESS in
correlated driven systems~\cite{Freericks2008,Freericks2008nds,Aoki2008,Aoki2009,Tsuji2010}.
Similar to the basic idea of static equilibrium DMFT~\cite{Georges1996},
it maps a correlated driven lattice to a driven impurity. To reach
a NESS, the driven lattice is coupled to an infinite equilibrium bath,
which can be fermionic~\cite{Aoki2009,Aoki2014} or bosonic~\cite{Lee2014di}.  In the following, we only consider a fermionic bath. The Hamiltonian of the total system is $H_{tot}\left(t\right)=H_{s}\left(t\right)+H_{b}+H_{sb}$,
where the subscript $s$ denotes system, $b$ denotes bath and $sb$
denotes system-bath coupling. The Hamiltonian of system $H_{s}\left(t\right)$
is time periodic: $H_{s}\left(t\right)=H_{s}\left(t+\mathcal{T}\right)$.
We will specify the system Hamiltonian in the following. It generally
consists of a kinetic term and interaction term. One can drive the
kinetic term or the interaction for different purposes of quantum
simulation. The bath is $H_{b}=\sum_{i,p}\left(\epsilon_{b,p}-\mu_{b}\right)b_{i,p}^{\dagger}b_{i,p}$
where $b_{i,p}^{\dagger}$($b_{i,p}$) creates (annihilates) a fermion
on site $i$ with quantum number $p$ (for example, spin or orbital)
in the bath with energy $\epsilon_{i,p}$, and $\mu_{b}$ is the chemical
potential of the bath. The system-bath coupling reads $H_{sb}=\sum_{i,p}V_{p}\left(c_{i}^{\dagger}b_{i,p}+b_{i,p}^{\dagger}c_{i}\right)$,
where $c_{i}^{\dagger}$ ($c_{i}$) creates (annihilates) a fermion
in the system, and $V_{p}$ is the hybridization between the system
and state $p$ of the bath. For a correlated system, one can drive the kinetic energy or the interaction. 

It can be shown~\cite{Aoki2009,Aoki2014} that the coupling to a
free-fermion bath leads to a correction $\Sigma_{diss}$ to the self-energy
of the lattice
\begin{equation}
\hat{G}_{loc}\left(\omega\right)=\sum_{\bm{k}}\left[\hat{G}_{0\bm{k}}^{-1}\left(\omega\right)-\hat{\Sigma}_{lat}\left(\omega\right)-\hat{\Sigma}_{diss}\left(\omega\right)\right]^{-1}\label{Gloc}
\end{equation}
where the Green's functions and self-energy are defined in the Floquet
space~\cite{Aoki2008}, i.e. a Fourier transform of $G\left(t,t^{\prime}\right)$
on the Keldysh contour, which is suitable for describing the NESS.
Every Green's function has three components $\hat{G}\left(\omega\right)=\left(\begin{array}{cc}
\hat{G}^{R}\left(\omega\right) & \hat{G}^{K}\left(\omega\right)\\
0 & \hat{G}^{A}\left(\omega\right)
\end{array}\right)$, with every component represented in Floquet space. In Eq.\eqref{Gloc} $\bm{k}$ is momentum. $\hat{G}_{loc}$ is the full
local Green's function. $\hat{G}_{0\bm{k}}\left(\omega\right)$ is
the non-interacting Green's function. $\hat{\Sigma}_{lat}\left(\omega\right)$
is the self-energy arising from the 2-particle interaction in the lattice model, which is obtained from the
impurity solver. For a free-fermion
bath, with a constant density of states approximation we have~\cite{Aoki2009} $\hat{\Sigma}_{diss}\left(\omega\right)=\left(\begin{array}{cc}
-\mathrm{i}\Gamma\mathbb{I} & -2\mathrm{i}\Gamma\bm{F}\left(\omega\right)\\
0 & \mathrm{i}\Gamma\mathbb{I}
\end{array}\right)$ where $\mathbb{I}$ is unit matrix, $\bm{F}_{mn}\left(\omega\right)\equiv\tanh\left(\frac{\omega+n\Omega}{2T}\right)\delta_{mn}$
and $m$ ($n$) is a Floquet index. $\Gamma$ is the damping rate
and $T$ is the bath temperature. We set $\Gamma=T=0.05$ in all our calculations in the following. The energy unit is the bare hopping of the system. The local self-energy is calculated
by using an impurity solver. The self-consistency loop is closed by
the equation for the dynamical mean field $\hat{\mathcal{G}}_{0}^{-1}\left(\omega\right)$
\begin{equation}
\hat{\mathcal{G}}_{0}^{-1}\left(\omega\right)=\hat{G}_{loc}^{-1}\left(\omega\right)+\hat{\Sigma}_{loc}\left(\omega\right).
\end{equation}

In our calculations we consider a Falicov-Kimball  interaction
\begin{equation}
H_{int}=U\sum_{i}c_{i}^{\dagger}c_{i}f_{i}^{\dagger}f_{i}
\end{equation}
where the $f$ atoms are localized. The impurity Green's function
is exactly given by~\cite{Freericks2008} 
\begin{equation}
\hat{G}\left(\omega\right)=w_{0}\hat{\mathcal{G}}_{0}\left(\omega\right)+w_{1}\hat{R}\left(\omega\right)\label{eq:kinsolver}
\end{equation}
where $\hat{R}\left(\omega\right)=\left(\hat{\mathcal{G}}_{0}^{-1}\left(\omega\right)-\hat{U}\left(\omega\right)\right)^{-1}$,
and $w_{1}$ is the probability that a site is being
occupied by immobile atoms with $w_0=1-w_1$. In our calculation we use $w_{0}=w_{1}=\frac{1}{2}$. 
As we show in the appendix~\ref{subsec:Impurity-solver}, for a driven
kinetic energy, $\hat{U}\left(\omega\right)$ is a diagonal matrix, while
for a driven interaction $\hat{U}\left(\omega\right)$ is replaced by a
matrix of the form $U+\delta U\cos\left(\Omega t\right)$ in Floquet
space, which is tri-diagonal. 

We would like to point out the relation between the Falicov-Kimball
model and a system with quenched binary disorder. The Falicov-Kimball model is
a generic correlated model, and it corresponds to a model with annealed disorder
in the equilibrium state~\cite{Eckstein2009}. For  the homogeneous case of half-filling with $w_{0}=w_{1}=\frac{1}{2}$,
if one uses the DMFT solution for the Falicov-Kimball model and the
coherent potential approximation~\cite{Yonezawa1973cpa,Elliott1974tpr} for a system with quenched binary disorder,
the two solutions are exactly the same. There is no such equivalence
for the general case~\cite{Eckstein2009,Brandt:1989aa}. 

We consider two different types of driving in an infinite-dimensional hypercubic lattice. 

\emph{ac driven kinetic energy}: We consider the Hamiltonian for a
shaken optical lattice in the lattice frame, 
\begin{equation}
H_{kin}\left(t\right)=-J\sum_{\left\langle ij\right\rangle }c_{i}^{\dagger}c_{j}+\sum_{i}\bm{E}\cdot\bm{R}_{i}\cos\left(\Omega t\right)c_{i}^{\dagger}c_{i}\label{eq:Hkin}
\end{equation}
where $J$ is the hopping
amplitude, and the force $\bm{E}=E\left(1,\cdots,1\right)$ is defined
in an infinite-dimensional hypercubic lattice. $\bm{R}_{i}$ is the lattice vector. This Hamiltonian can also be used to describe a system in a spatially
uniform but time dependent electric field $\bm{E}\cos\Omega t$. 
With the unitary transformation $V\left(t\right)=e^{-\mathrm{i}\frac{1}{\Omega}\sin\left(\Omega t\right)\sum_{i}\bm{E}\cdot\bm{R}_{i}c_{i}^{\dagger}c_{i}}$,
$\tilde{H}_{kin}=V^{\dagger}H_{kin}V-\mathrm{i}V^{\dagger}\partial_{t}V$. The purpose of this transformation is to remove the
time-dependent term of the Hamiltonian in Eq.~\eqref{eq:Hkin}. The quantum state
is transformed by $\left|\tilde{\psi}\right\rangle =V^{\dagger}\left(t\right)\left|\psi\right\rangle $.
This transformation is equivalent to a lattice momentum shift of the
quantum state, namely $V^{\dagger}\left(t\right)=e^{\mathrm{i}\frac{1}{\Omega}\sin\left(\Omega t\right)\sum_{i}\bm{E}\cdot\bm{R}_{i}c_{i}^{\dagger}c_{i}}$.
The force in the reference frame of lattice is $\bm{F}=\bm{E}\cos\left(\Omega t\right)$,
therefore $\frac{1}{\Omega}\sin\left(\Omega t\right)\bm{E}=\int dt\bm{F}\left(t\right)$
is the momentum.
Then 
\begin{equation}
\tilde{H}_{kin}\left(t\right)=-J\sum_{\left\langle ij\right\rangle }e^{\mathrm{i}\frac{\sin\left(\Omega t\right)}{\Omega}\bm{E}\cdot\left(\bm{R}_{i}-\bm{R}_{j}\right)}c_{i}^{\dagger}c_{j}.\label{eq:hK}
\end{equation}
 In momentum space, it has the form~\cite{Aoki2008}
\begin{equation}
\tilde{H}_{kin}\left(t\right)=\sum_{\bm{k}}\epsilon_{\bm{k}-\bm{A}\left(t\right)}c_{\bm{k}}^{\dagger}c_{\bm{k}}\label{eq:ac}
\end{equation}
where $A\left(t\right)=\frac{E}{\Omega}\sin\left(\Omega t\right)$
and $\bm{A}\left(t\right)=A\left(t\right)\left(1,\cdots,1\right)$. We choose the lattice constant as length unit.
Models of this type are of general interest. The kinetic part of a
shaken optical lattice can be cast in the form of Eq.~(\ref{eq:ac})~\cite{Jotzu:2014fk,Jotzu2015a,Gorg:2017aa}.
With proper consideration of lattice symmetry, this model can be used
to study shaken optical lattices with different driving protocols.

\emph{ac driven interaction}: Following the seminal proposals in Refs.~\cite{Rapp2012ulg,Greschner2014dds}
and recent experimental progress~\cite{Meinert2016fec}, we study
NESS in a system with driven interactions, which is achieved by using
a Feshbach resonance induced by an oscillating magnetic field~\cite{Rapp2012ulg,Meinert2016fec}.
We consider an ac-driven interaction 
\begin{equation}
H_{int}\left(t\right)=\left(U+\delta U\cos\left(\Omega t\right)\right)\sum_{i}c_{i}^{\dagger}c_{i}f_{i}^{\dagger}f_{i}.
\end{equation}

\section{\label{sec:Physical-quantities}Physical quantities}

For the Falicov-Kimball model, we consider the double occupancy, $D\left(t\right)=-\mathrm{i}w_{1}R\left(t\right)$~\cite{Eckstein2009}. The physical meaning of $D(t)$ is the probability of a mobile atom and an immobile atom occupying the same site.
For a NESS, we have $D=\frac{1}{\mathcal{T}}\int_{0}^{\mathcal{T}}dtD\left(t\right)=w_{1}\sum_{m}\int_{0}^{\Omega}\frac{d\omega}{2\pi}\mathrm{Im}\left[R_{mm}^{<}\left(\omega\right)\right]$.
The second important quantity is the fraction of atoms excited to
the upper Mott band. To show that the final converged DMFT solution
is a NESS, we also need to calculate the work $W$ done by the driving
field, and the energy dissipation rate $I$ into the bath. Details
of deriving $W$ and $I$ can be found in Ref.~\cite{Tsuji2010}.
We briefly outline the procedure. The rate of change of the internal
energy of the system is $\frac{dE_{s}\left(t\right)}{dt}=\left\langle \frac{\partial H_{s}\left(t\right)}{\partial t}\right\rangle +\mathrm{i}\left\langle \left[H_{tot},H_{s}\left(t\right)\right]\right\rangle \equiv W-I$.
For a NESS, $I=-\mathrm{i}\left\langle \left[H_{tot},H_{s}\right]\right\rangle =\mathrm{i}\left\langle \left[H_{tot},H_{b}\right]\right\rangle$. 
In forms of Floquet Green's functions, one finds that~\cite{Tsuji2010}
\begin{align}
I & =-\mathrm{i}\Gamma\sum_{\bm{k},n}\int_{0}^{\Omega}\frac{d\omega}{2\pi}(\omega+n\Omega)\Biggl\{ G_{\bm{k},nn}\left(\omega\right)\nonumber \\
 & -F_{nn}\left(\omega\right)\left[G_{\bm{k},nn}^{R}\left(\omega\right)-G_{\bm{k},nn}^{A}\left(\omega\right)\right]\Biggr\}.\label{eq:I}
\end{align}
For the case of driven kinetic energy, one can also easily derive
$W=\frac{E}{2}\sum_{mn,\bm{k}}\left(v_{\bm{k}}^{n+1,m}+v_{\bm{k}}^{n-1,m}\right)\int_{0}^{\Omega}\frac{d\omega}{2\pi}G_{\bm{k},mn}^{<}\left(\omega\right)$,
with the velocity $v_{\bm{k}}^{mn}=\frac{1}{\mathcal{T}}\int_{0}^{\mathcal{T}}dte^{\mathrm{i}\left(m-n\right)\Omega t}v_{\bm{k}-\bm{A}\left(t\right)}$.
For a driven interaction, one can show that 
\begin{align}
W= & \mathrm{i}\Omega\delta U\frac{1}{\mathcal{T}}\int_{0}^{\mathcal{T}}dt\sin\left(\Omega t\right)\left(w_{1}R^{<}\left(t,t\right)-\frac{1}{2}G^{<}\left(t,t\right)\right)\nonumber \\
= & \frac{\Omega\delta U}{2}\sum_{mn}\int_{0}^{\Omega}\frac{d\omega}{2\pi}\left(w_{1}R_{mn}^{<}\left(\omega\right)-\frac{G_{mn}^{<}\left(\omega\right)}{2}\right)\nonumber \\
 & \times\left(\delta_{m,n+1}-\delta_{m+1,n}\right),\label{eq:UdriveW}
\end{align}
where the contribution of $G_{nm}^{<}\left(\omega\right)$ part is
from the chemical potential (See Appendix~\ref{subsec:Impurity-solver}).
In the following, we will discuss our numerical simulation results
for these physical quantities.

\section{\label{sec:Results-and-discussions}Results and discussions}

\subsection{AC-driven kinetic energy}

In this section, we present nonequilibrium and resonant properties
of the NESS for a Falicov-Kimball model with driven kinetic energy.
As we will show in the following, the NESS has well-defined Mott
bands in the strongly correlated regime, which are renormalized by
the driving. The interplay between interaction $U$ and driving frequency
$\Omega$ leads to resonances and a dramatic change of occupancies
of different bands. 

\subsubsection{Double occupancy and resonance}

\begin{figure}[h]
\includegraphics[scale=0.7]{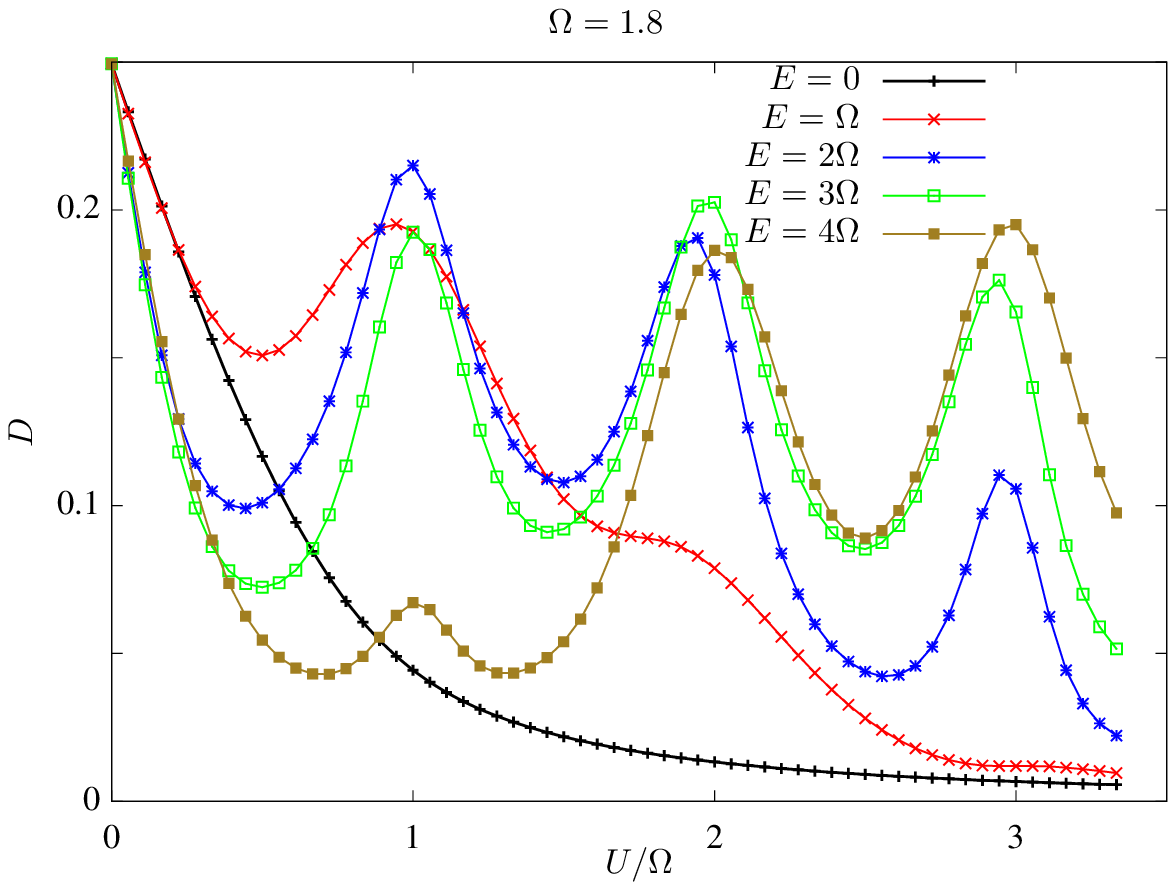}

\includegraphics[scale=0.7]{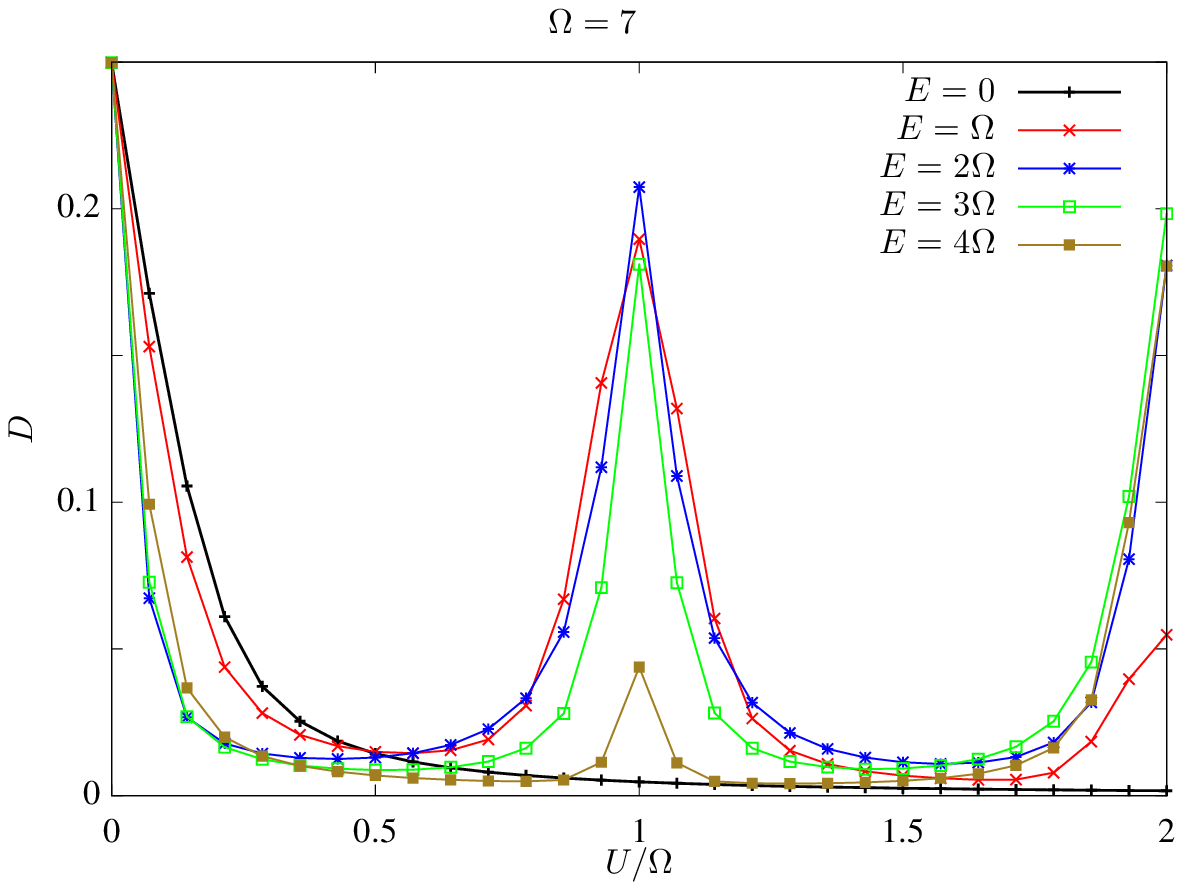}

\caption{\label{fig:doubleome}Double occupancy for the Falicov-Kimball
model with ac-driven kinetic energy.}
\end{figure}

In the double occupancy in Fig.~\ref{fig:doubleome}, compared to
the non-driven case $E=0$, we clearly observe resonances at interaction
values $U=n\Omega$ with integer $n\ge1$, which is corresponding
to $n$-photon absorption. This resonant behavior, which occurs for
both choices of $\Omega$ in Fig.~\ref{fig:doubleome} shows nonequilibrium
properties of the NESS. The Mott insulator transition for the Falicov-Kimball
model occurs at $U\approx1.4$, where we choose the bare hopping as
the energy unit, therefore $\Omega=1.8$ is not a small energy scale.
As a result, in the strongly correlated regime, one can expect resonant tunneling
when $U=n\Omega$. From Eq.~(\ref{eq:hK}), $\tilde{H}_{kin}\left(t\right)=-J\sum_{\left\langle ij\right\rangle }\sum_{n=-\infty}^{\infty}\mathcal{J}_{n}\left(\frac{\bm{E}\cdot\left(\bm{R}_{i}-\bm{R}_{j}\right)}{\Omega}\right)e^{\mathrm{i}n\Omega t}c_{i}^{\dagger}c_{j}$
where $\mathcal{J}_{n}\left(\frac{\bm{E}\cdot\left(\bm{R}_{i}-\bm{R}_{j}\right)}{\Omega}\right)$
is the $n$-th Bessel function of the first kind. One can see that
a $n\Omega$ resonance can be of first order, because the driving
$e^{in\Omega t}$ with frequency $n\Omega$ appears in $\tilde{H}_{kin}$.
We notice that the resonance behavior of double occupancy has been
observed in a recent experiment~\cite{Gorg:2017aa} for the Hubbard
model. The Falicov-Kimball model, which we investigate here as a limiting
case of the Hubbard model, can thus capture the relevant single-band
physics of the system.

\subsubsection{Resonance and Fermi's golden rule}

\begin{figure}[h]
\includegraphics[scale=0.7]{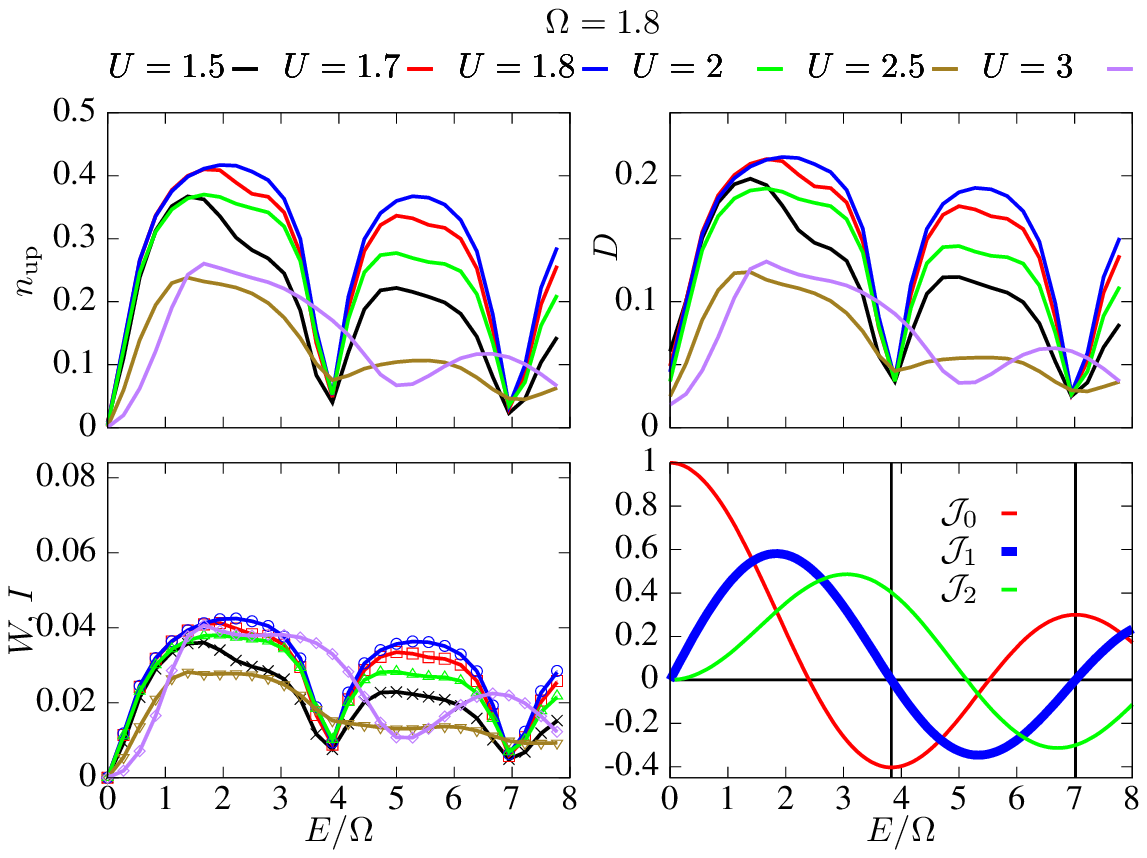}

\includegraphics[scale=0.7]{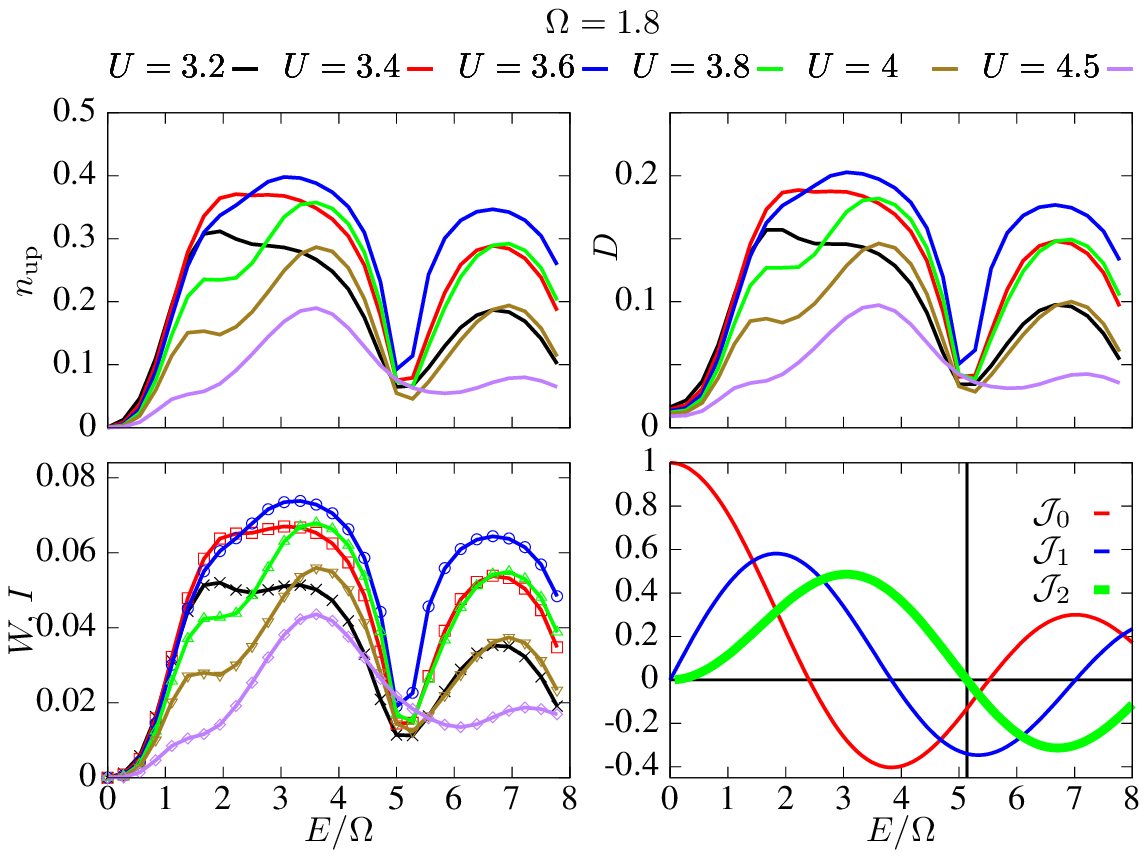}

\caption{\label{fig:FKheating}For the Falicov-Kimball
model with ac-driven kinetic energy, we
show the fraction of atoms excited to the upper band $n_{\mathrm{up}}$,
double occupancy $D$, work $W$ (dots), energy dissipation $I$ (lines)
and Bessel function of the first kind $\mathcal{J}_{l}\left(\frac{E}{\Omega}\right)$.
At the one-photon resonance (upper panel), the oscillation is qualitatively
described by $\left|\mathcal{J}_{1}\left(\frac{E}{\Omega}\right)\right|$,
while the oscillation at the two-photon resonance qualitatively coincides
with $\left|\mathcal{J}_{2}\left(\frac{E}{\Omega}\right)\right|$. }
\end{figure}

In Fig.~\ref{fig:FKheating}, we show (near-) resonant behavior of
an ac-driven Falicov-Kimball model for values of $U$ close to $\Omega$
or $2\Omega$. The fraction of atoms in the higher Mott band $n_{\mathrm{up}}$,
and the double occupancy $D$ are different from their equilibrium
values when the driving amplitude $E$ is finite. We observe that
$W=I$ in the full parameter range which we have explored. This implies
that the system reaches the NESS. We clearly observe that $n_{\mathrm{up}}$,
$D$, $W$ and $I$ have similar behavior as a function of $\frac{E}{\Omega}$,
which can be understood as follows. In the strongly correlated regime, if atoms
are excited to the higher Mott band, the double occupancy will
increase, since most sites are singly occupied. Correspondingly, this
means that the system absorbs energy from the driving fields. To keep
it in the NESS, more energy should therefore be dissipated to the
bath. 

In the upper panel of Fig.~\ref{fig:FKheating}, for $U$ close to
$\Omega$, we observe that the behavior of these observables is qualitatively
similar to $\left|\mathcal{J}_{1}\left(\frac{E}{\Omega}\right)\right|$,
while in the lower panel, for $U$ close to $2\Omega$, it is similar
to $\left|\mathcal{J}_{2}\left(\frac{E}{\Omega}\right)\right|$. It
can be qualitatively described by Fermi's golden rule. In the strongly
correlated regime, the kinetic part can be treated as a perturbation
relative to the interaction term. The transition amplitude between
initial state $\left|I\right\rangle $ and final state $\left|F\right\rangle $
is~\cite{Shirley1965sse,Bilitewski2015} 
\[
A\left(I\rightarrow F,t\right)=\frac{-\mathrm{i}}{\hbar}\int_{0}^{t}dt^{\prime}e^{-\mathrm{i}\left(\mathcal{E}_{I}-\mathcal{E}_{F}\right)t^{\prime}}\left\langle \phi_{F}\right|\tilde{H}_{kin}\left(t^{\prime}\right)\left|\phi_{I}\right\rangle 
\]
where the kinetic part is $\tilde{H}_{kin}\left(t\right)=\sum_{\bm{k}}\epsilon_{\bm{k}-\bm{A}\left(t\right)}c_{\bm{k}}^{\dagger}c_{\bm{k}}$
with $\epsilon_{\bm{k}-\bm{A}\left(t\right)}=\epsilon_{\bm{k}}\cos\left(\frac{E}{\Omega}\sin\left(\Omega t\right)\right)+\bar{\epsilon}_{\bm{k}}\sin\left(\frac{E}{\Omega}\sin\left(\Omega t\right)\right)$,
and where $\epsilon_{\bm{k}}=-2J\sum_{i=1}^{d}\cos k_{i}$ and $\bar{\epsilon}_{\bm{k}}=-2J\sum_{i=1}^{d}\sin k_{i}$.
The transition probability $\gamma_{I\rightarrow F}=\lim_{t\rightarrow\infty}\frac{\left|A\left(I\rightarrow F,t\right)\right|^{2}}{t}$
takes the form
\begin{align}
\gamma_{I\rightarrow F} & =\sum_{\bm{k},\bm{k}^{\prime}}n_{\bm{k}}^{FI}n_{\bm{k}^{\prime}}^{IF}\nonumber \\
\times & \biggl(\epsilon_{\bm{k}}\epsilon_{\bm{k}^{\prime}}\sum_{l=even}\delta\left(\mathcal{E}_{I}-\mathcal{E}_{F}+l\Omega\right)\left|\mathcal{J}_{l}\left(z\right)\right|^{2}\nonumber \\
+ & \bar{\epsilon}_{\bm{k}}\bar{\epsilon}_{\bm{k}^{\prime}}\sum_{l=odd}\delta\left(\mathcal{E}_{I}-\mathcal{E}_{F}+l\Omega\right)\left|\mathcal{J}_{l}\left(z\right)\right|^{2}\biggr)\label{eq:FGR}
\end{align}
with $n_{\bm{k}}^{IF}=\left\langle I\right|c_{\bm{k}}^{\dagger}c_{\bm{k}}\left|F\right\rangle $
and $z=\frac{E}{\Omega}$. In the strongly correlated regime of an ac-driven
Falicov-Kimball model, $\mathcal{E}_{\mathrm{I}}$ and $\mathcal{E}_{\mathrm{F}}$
are corresponding to the lower and upper Mott bands. From Eq.~(\ref{eq:FGR})
we conclude that in an insulating state a $n\Omega$ ($n\ge1$) resonance
would be qualitatively described by $\left|\mathcal{J}_{n}\left(\frac{E}{\Omega}\right)\right|$. 

Alternatively, we can study the driven, strongly correlated regime
using the Schrieffer-Wolff transformation~\cite{bukov2015,Bukov2016sf}.
With a further rotation $V^{\prime}\left(t\right)=\exp\left[-\mathrm{i}Ut\sum_{i}c_{i}^{\dagger}c_{i}f_{i}^{\dagger}f_{i}\right]$
of the Hamiltonian $H=\tilde{H}_{kin}+H_{int}$, we have $\tilde{H}^{rot}=-J\sum_{l}\sum_{\left\langle ij\right\rangle }\mathcal{J}_{l}\left(z_{ij}\right)e^{\mathrm{i}l\Omega t+\mathrm{i}Ut\left(n_{fi}-n_{fj}\right)}c_{i}^{\dagger}c_{j}$.
For the resonant case where $U=m\Omega$ with integer $m$, we have
the effective Hamiltonian from the high-frequency expansion~\cite{Bukov2016sf}, 

\begin{equation}
H_{eff}^{\left(0\right)}=-J\sum_{\left\langle ij\right\rangle }\left(\mathcal{J}_{0}\left(z_{ij}\right)\hat{g}_{f,ij}+\mathcal{J}_{m}\left(z_{ij}\right)\hat{h}_{f,ij}\right)c_{i}^{\dagger}c_{j}\label{eq:heff}
\end{equation}
where $z_{ij}=\frac{\bm{E}\cdot\left(\bm{R}_{i}-\bm{R}_{j}\right)}{\Omega}$
is limited to nearest neighbors, $\hat{g}_{f,ij}=\left(1-n_{fi}\right)\left(1-n_{fj}\right)+n_{fi}n_{fj}$
and $\hat{h}_{f,ij}=(-1)^m n_{fi}\left(1-n_{fj}\right)+\left(1-n_{fi}\right)n_{fj}$
with $n_{fi\left(j\right)}=f_{i\left(j\right)}^{\dagger}f_{i\left(j\right)}$.
In Eq.~(\ref{eq:heff}), clearly the second term is introduced due
to the resonance. We observe that with the same amplitude $\mathcal{J}_{m}\left(z_{ij}\right)$
the correlated hopping $c_{i}^{\dagger}c_{j}n_{fi}\left(1-n_{fj}\right)$
creates a double occupancy on site $i$, and that $c_{i}^{\dagger}c_{j}\left(1-n_{fi}\right)n_{fj}$
destroys a double occupancy. This explains why in Fig.~\ref{fig:FKheating}
the double occupancy tends to follow the Bessel functions. It is consistent
with our understanding from Fermi's golden rule. 

In Ref.~\cite{Jordens:2008aa}, the energy gap of the Mott insulator
of a Fermi-Hubbard model is determined by a distinct peak in the double
occupancy when $U$ matches the modulation frequency. It can be understood
as one-photon resonant tunneling, as we show in Figs.~\ref{fig:doubleome}
and~\ref{fig:FKheating}.

In the strongly correlated regime, an occupation of the upper Mott
band dramatically changes the effective distribution $f_{\mathrm{eff}}\left(\omega^{\prime}\right)=\frac{N\left(\omega^{\prime}\right)}{A\left(\omega^{\prime}\right)}$
where $A\left(\omega^{\prime}\right)=-\frac{1}{\pi}\mathrm{Im}G^{R}_{nn}\left(\omega\right)$
and $N\left(\omega^{\prime}\right)=\frac{1}{2\pi}\mathrm{Im}G^{<}_{nn}\left(\omega\right)$. $\omega\in\left[-\frac{\Omega}{2},\frac{\Omega}{2}\right)$, and $\omega^{\prime}=\omega+n\Omega$.
For the static case with half-filling, one has $N\left(\omega^{\prime}>0\right)=0$
while for the driven case $N\left(\omega^{\prime}>0\right)$ is finite,
therefore $f_{\mathrm{eff}}\left(\omega^{\prime}\right)$ is very
different from a Fermi-Dirac distribution. This is a characteristic
of the NESS. For a static case, there are two Mott-bands separated
by $U$. With driving turned on, there are quasi-energy bands with
$n\Omega$ ($n\ge1$ integer) energy difference from the bands
of the static case. Excitations to the higher quasi-energy bands need
assistance of a photon with frequency $n\Omega$. By occupation of the higher Mott
band, the system absorbs energy from driving and dissipates it to
the bath, so the energy dissipation rate $I$ is finite. From Eq.~(\ref{eq:I}) we
see that a nonzero $I$ implies a violation of the fluctuation-dissipation
theorem~\cite{Tsuji2010}.

\begin{figure}[h]
\includegraphics[scale=0.7]{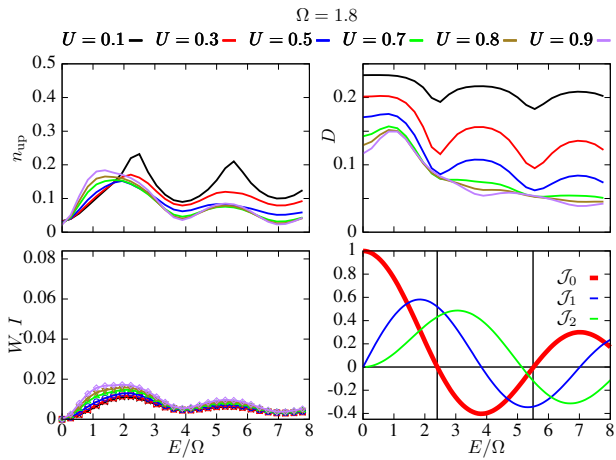}

\caption{\label{fig:J0}For  the Falicov-Kimball
model with ac-driven kinetic energy, the fraction
of atoms excited to the upper band $n_{\mathrm{up}}$, double occupancy
$D$, work $W$ (dots), energy dissipation $I$ (lines) and Bessel
function of the first kind $\mathcal{J}_{l}\left(\frac{E}{\Omega}\right)$
are shown in the weakly interacting regime.}
\end{figure}

We also study the nonequilibrium behavior in the weakly-interacting
regime. In Fig.~\ref{fig:J0}, we find the double occupancy to be
qualitatively controlled by $\left|\mathcal{J}_{0}\left(\frac{E}{\Omega}\right)\right|$
for weak $U$. It shows that the system is described by the effective
Hamiltonian and in a (correlated) ``metallic'' state. A resonance
is possible, as shown in the plot for $W$ and $I$ in Fig.~\ref{fig:J0},
if the driving frequency can match the band-width. The minima of $W$
and $I$ are at the nodes of $\mathcal{J}_{1}\left(z\right)=0$ as
before. This indicates that only the process of single-photon absorption
occurs. However, the resonance is greatly suppressed. 

To sum up, we have shown that in an ac-driven Falicov-Kimball model,
driving frequency, driving amplitude and interaction strength are
important parameters for controlling physical properties of the NESS.
With a careful choice of these parameters, one can tune the NESS of
the driven system very close to an effective equilibrium system. With
good control of resonant tunneling, one can induce different populations
of different Floquet bands and achieve the goal of ``photo-doping''. 

\subsection{Driven Interaction}

In this section, we analyze the behavior of double occupancy due to
periodic modulation of the interaction. We explain it by analytical
calculation and using the information from spectral functions. We
show that the NESS is reached within the framework of Floquet DMFT. 

\begin{figure}[h]
\includegraphics[scale=0.6]{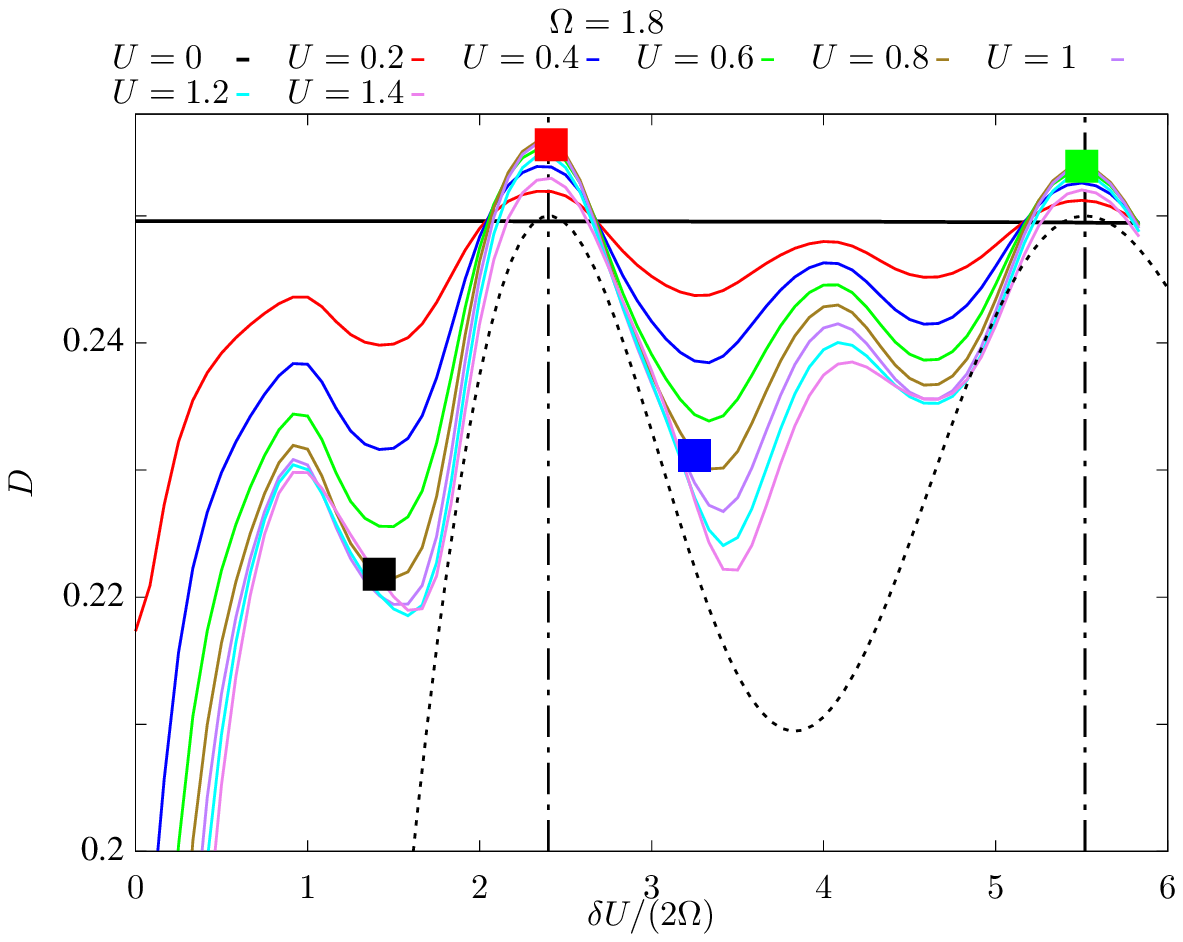}

\includegraphics[scale=0.6]{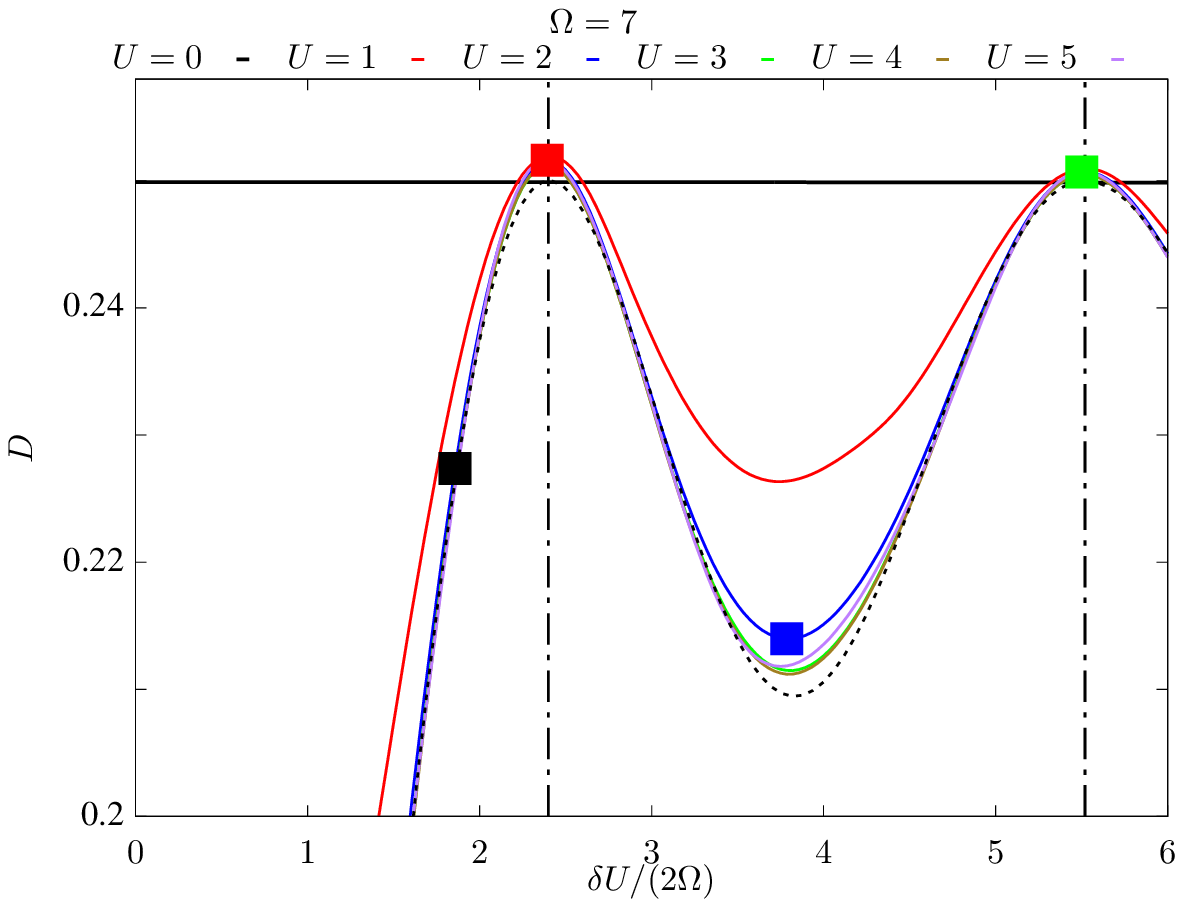}

\caption{\label{fig:D_driving_amplitudes}Double occupancy versus driving amplitude
$\delta U$ for different values of $U$ and the driving frequency
$\Omega$.}
\end{figure}

In Fig.~\ref{fig:D_driving_amplitudes}, we plot the modulation of
double occupancy $D$ versus driving amplitude for different interactions.
We note the following observations. (i) When $U=0$, $D$ is not modulated
by $\delta U$. This is due to the symmetry in the system Hamiltonian~\cite{Liberto2014qsc}.
(ii) When $U\neq0$, we find that $D$ is modulated as a function
of $\delta U$. (iii) Close to the position of the nodes ($\frac{\delta U}{2\Omega}=2.4,\;5.5,$
$\cdots$) of $\mathcal{J}_{0}\left(\frac{\delta U}{2\Omega}\right)$,
we find that $D>\frac{1}{4}$ when $\delta U\gg U$. In the following,
we explain (ii) and (iii) in order.

\begin{figure}[h]
\includegraphics{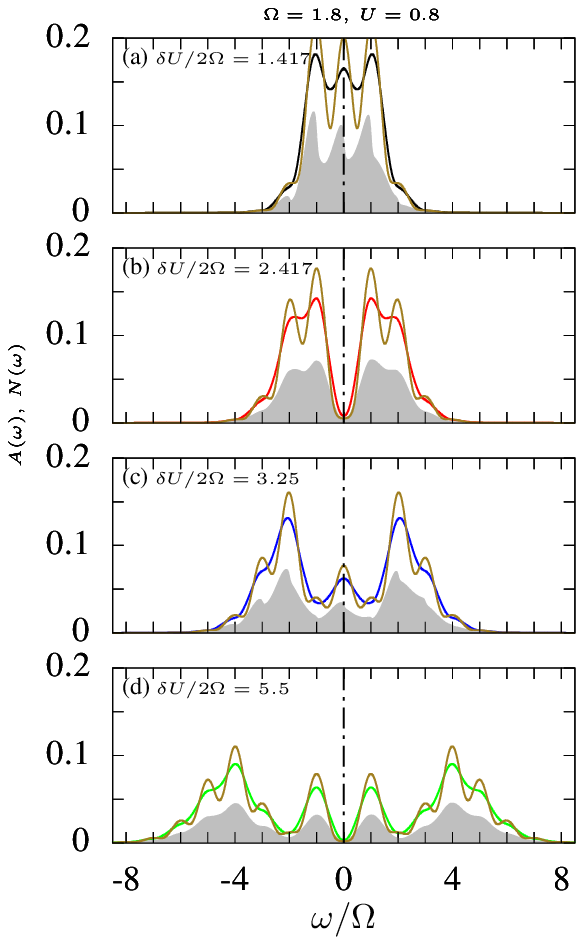}

\includegraphics{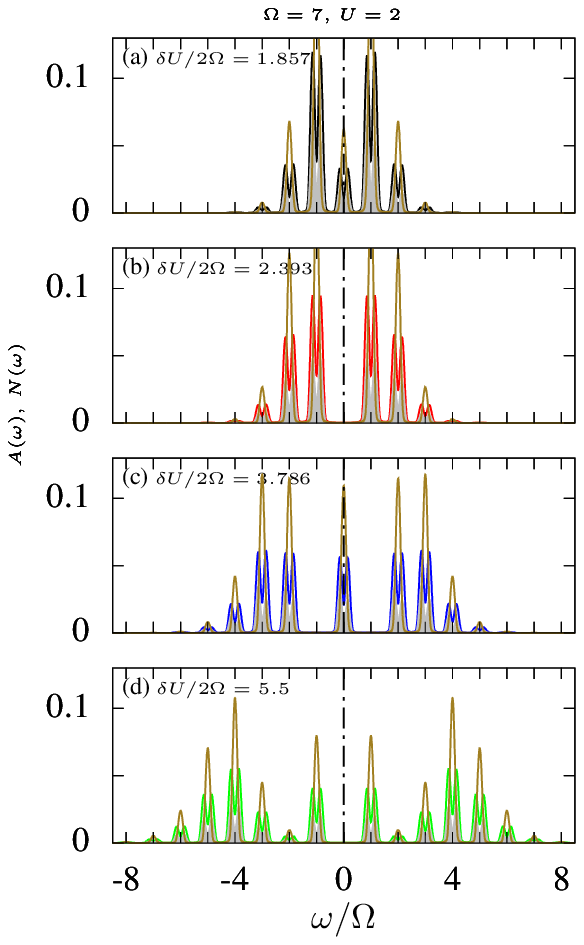}

\caption{\label{fig:Spectral_1d8_and_7}Spectral function $A\left(\omega\right)$
and occupied density of states $N\left(\omega\right)$ for points
with same color indicated in Fig.~\ref{fig:D_driving_amplitudes}.
The olive lines are corresponding to $U=0$.}
\end{figure}

We first explain that the modulation of $D$ can be qualitatively
described by $\mathcal{J}_{0}^{2}\left(\frac{\delta U}{2\Omega}\right)$.
As shown in Sec.~\ref{subsec:FGFM}, the non-interacting retarded
Floquet Green's function for a system with time-periodically modulated
interactions is
\begin{align}
G_{mn}^{R\left(0\right)}\left(\omega\right) & =\sum_{l}\mathcal{J}_{l-m}\left(\frac{\delta U}{2\Omega}\right)\mathcal{J}_{l-n}\left(\frac{\delta U}{2\Omega}\right)\nonumber \\
 & \times\left(-\mathrm{i}\sqrt{\pi}e^{-\xi_{l}^{2}}\mathrm{erfc}\left(-\mathrm{i}\xi_{l}\right)\right),
\end{align}
where $\xi_{l}=\omega+l\Omega+\frac{U}{2}+\mathrm{i}\Gamma$. One
can therefore expect that the dependence of the diagonal elements
of the interacting Green's function $\hat{G}^{R}\left(\omega\right)$ on
$\frac{\delta U}{2\Omega}$ are described by a function related to
the Bessel function $\mathcal{J}_{l}\left(\frac{\delta U}{2\Omega}\right)$.
One further obtains $\hat{G}^{<}\left(\omega\right)=(\hat{G}^{R}\left(-2i\Gamma\hat{\bm{F}}+\hat{\Sigma}_{lat}^{K}\right)\hat{G}^{A}-\hat{G}^{R}+\hat{G}^{A})/2$
where $F_{mn}=F\left(\omega+n\Omega\right)\delta_{mn}$. From the
impurity solver, we have $w_{1}\hat{R}\left(\omega\right)=\hat{G}\left(\omega\right)-w_{0}\hat{\mathcal{G}}_{0}\left(\omega\right)$.
Since the double occupancy is given by $D=w_{1}\sum_{m}\int_{0}^{\Omega}\frac{d\omega}{2\pi}\mathrm{Im}\left[R_{mm}^{<}\left(\omega\right)\right]$,
we conclude that $D$ is modulated by a function related to the Bessel
function. Based on this observation, we plot $g\left(\frac{\delta U}{2\Omega}\right)=-\frac{1}{4}\mathcal{J}_{0}^{2}\left(\frac{\delta U}{2\Omega}\right)+\frac{1}{4}$
(black dashed line) in Fig.~\ref{fig:D_driving_amplitudes}. We find
that for the case of high driving frequency $D$ is well consistent
with $g\left(\frac{\delta U}{2\Omega}\right)$. It is not easy to
find an analytical formula for the interacting case, because all quantities
involved are matrices. We present in Fig.~\ref{fig:Spectral_1d8_and_7}
spectral functions for points of same color indicated in Fig.~\ref{fig:D_driving_amplitudes}.
We see that the density of states (DOS) around $\omega=0$ is tiny
when $\frac{\delta U}{2\Omega}\approx2.4,\,5.5$ compared to other
values, which indicates that the DOS at $\omega=0$ is modulated by
$\mathcal{J}_{0}^{2}\left(\frac{\delta U}{2\Omega}\right)$. A change
in double occupancy $D$ involves excitations over the Fermi level
at $\omega=0$. The small value of the DOS at $\omega=0$ makes high-order
excitations very difficult because they require intermediate states
close to $\omega=0$. Note that for this reason the gap around $\omega=0$
has approximately the width $2\Omega$ (see Fig.~\ref{fig:Spectral_1d8_and_7}).
This is the reason why $D$ can be affected by a modulation of the
DOS around $\omega=0$. 

We now explain why $D>\frac{1}{4}$ close to the nodes of $\mathcal{J}_{0}\left(\frac{\delta U}{2\Omega}\right)$
when $\delta U\gg U$. Using nonequilibrium DMFT, it has been found
that $D\left(t\right)>\frac{1}{4}$ in a transient dynamics of a (driven)
Hubbard model because of band flipping~\cite{Tsuji2011,Tsuji2012},
or a favoring of holon-doublon process due to dynamical localization~\cite{MendozaArenas2017}.
In our case, the situation is different because our system is driven
in a different way, and we focus on the final NESS. For $\Omega=7$
and $U=2$ in Fig.~\ref{fig:D_driving_amplitudes}, we look at the
change of $D$ with increasing driving amplitude. For finite $U$,
$D$ is small when $\delta U=0$. It is greatly enhanced because of
photon-assisted tunneling with increasing $\delta U$. $D$ saturates
close to the $\frac{\delta U}{2\Omega}\approx2.4$, i.e. at the first
node of $\mathcal{J}_{0}\left(\frac{\delta U}{2\Omega}\right)=0$,
where the opening of a gap of the order of $2\Omega$ makes photon-assisted
tunneling difficult (see Fig.~\ref{fig:Spectral_1d8_and_7}). For
$\Omega=1.8$, the change of $D$ is more complex because high-order
processes are involved. Therefore, the role of photon-assisted tunneling
is very important for the enhancement of $D$. At the nodes of $\mathcal{J}_{0}\left(\frac{\delta U}{2\Omega}\right)=0$,
photon-assisted tunneling is possible to favor $D>\frac{1}{4}$ when
$U$ is finite. A finite $U$ has two effects: it makes creation of
doublon difficult, and it increases the band width to facilitate photon
assisted tunneling over $2\Omega$ gap, which will further increase
$D$. Figure~\ref{fig:Spectral_1d8_and_7} shows that the band width
is enhanced by finite $U$ compared to the case $U=0$ (see olive
lines in Fig.~\ref{fig:Spectral_1d8_and_7}). Our calculation shows
that the second effect dominates in the regime $\delta U\gg U,\Omega$. 

\begin{figure}[h]

\includegraphics[scale=0.55]{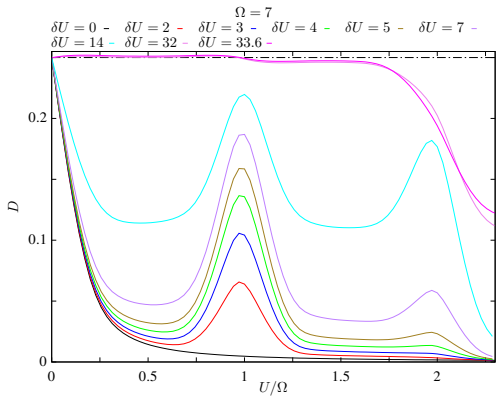}

\caption{\label{fig:D-Udrive}Double occupancy versus $U$ for different driving
amplitudes with driving frequency $\Omega=7$.}
\end{figure}

In Fig.~\ref{fig:D-Udrive}, the double occupancy is shown for different
driving amplitudes and we see that the system is in the non-equilibrium state. (i)
Compared to the non-driven case, the double occupancy $D$ is dramatically
changed. For
this high driving frequency ($\Omega=7$), we can see clear resonance
peaks when $U=n\Omega$. As we have shown for the case of a driven
kinetic energy, a well-defined resonant peak is corresponding to two
well-separated Mott bands. (ii) For high-driving frequency and
$\delta U=33.6$, i.e., $\frac{\delta U}{2\Omega}=2.4$, $D$ is close
to 0.25. This is a reflection of the behavior of the double occupancy
around the nodes of $\mathcal{J}_{0}\left(\frac{\delta U}{2\Omega}\right)$,
which we observed in Fig.~\ref{fig:D_driving_amplitudes}. It will
not change dramatically until $U$ becomes the dominant energy scale. 

\begin{figure}[h]

\includegraphics[scale=0.5]{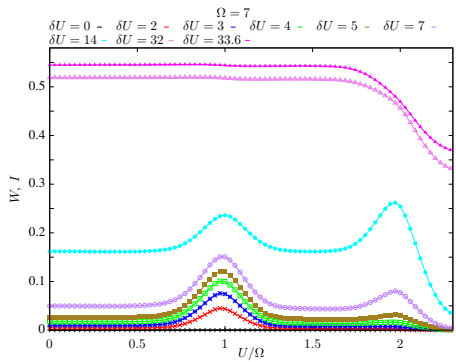}

\caption{\label{fig:WandI-Udrive}Work done by the driving field $W$ (dots),
and energy dissipation to the bath $I$ (lines) as a function of $U$
for different driving amplitudes with driving frequency $\Omega=7$. }
\end{figure}

Finally, in Fig.~\ref{fig:WandI-Udrive}, we clearly show that $W=I$
for the parameters in Fig.~\ref{fig:D-Udrive}, which indicates that
the NESS is reached. The work $W$ is defined as in Eq.~(\ref{eq:UdriveW}).
When the driving amplitude $\delta U=0$, there is no energy dissipation
to the bath. When $U\neq n\Omega$, no resonant tunneling is possible
and the dissipation is suppressed to relatively small values compared with those of resonant cases. This provides important
information about parameter regions where the NESS is close to an
equilibrium state.

To sum up, for time-periodically driven interactions, we show that
relevant physical quantities are dramatically changed compared to
the non-driven case. We find that the double occupancy is modulated
as $\mathcal{J}_{0}^{2}\left(\frac{\delta U}{2\Omega}\right)$, and
that it can exceed the value $\frac{1}{4}$ close to nodes of $\mathcal{J}_{0}\left(\frac{\delta U}{2\Omega}\right)$
when $\delta U\gg U$. We furthermore show that the NESS is reached
throughout the parameter region which we have explored. 

\section{\label{sec:Conclusions}Conclusions}

In conclusion, we have systematically studied the NESS in a driven
Falicov-Kimball model, with driven kinetic energy or driven interaction.
(i) For an ac-driven kinetic energy, we show that the relevant physical
quantities in the strongly correlated regime around the resonance
can be qualitatively described by Bessel functions. We explain this
phenomenon by Fermi's golden rule, and by the Schrieffer-Wolff transformation
of the driven Hamiltonian. These two understandings are consistent.
Resonant tunneling is an efficient way to control the occupancy of
Floquet bands and it can be used to achieve ``photo-doping''. (ii)
For driven interactions, we find that the double occupancy is dramatically
modulated as a function of the driving amplitude. We find that close
to the nodes of $\mathcal{J}_{0}\left(\frac{\delta U}{2\Omega}\right)$
the double occupancy can be larger than $\frac{1}{4}$ when $\delta U\gg U$,
and we argue that this can be understood by taking into account the
special role played by the DOS close to $\omega=0$, as well as the
role of a small finite positive $U$. We demonstrate that the NESS
is reached in both cases i) and ii).

In the future, we will investigate systems with double modulation,
i.e. with simultaneously driven kinetic energy and driven interaction.
As shown in Refs.~\cite{Greschner2014euh,Greschner2014dds}, one-dimensional
strongly correlated systems with double modulation can have exotic
and interesting physics. This will motivate studies of high-dimensional
systems with double modulation using the non-perturbative method of
Floquet DMFT. 
\begin{acknowledgments}
The authors acknowledge useful discussions or communications with
K. Le Hur, M. Eckstein, T. Oka, and N. Tsuji. This work was supported
by the Deutsche Forschungsgemeinschaft via DFG FOR 2414 and the high-performance
computing center LOEWE-CSC.
\end{acknowledgments}

\section{Appendix}

\subsection{\label{subsec:Impurity-solver}Impurity solver for the Falicov-Kimball
model with driven interactions}

We derive the impurity solver for the Falicov-Kimball model with driven
interactions. In the time domain, the impurity solver for the Falicov-Kimball
model can be written as~\cite{Aoki2014}
\begin{equation}
G\left(t,t^{\prime}\right)=w_{0}Q\left(t,t^{\prime}\right)+w_{1}R\left(t,t^{\prime}\right)
\end{equation}
where $Q\left(t,t^{\prime}\right)$ and $R\left(t,t^{\prime}\right)$
are defined by the following equations, 

\begin{align}
\left[\mathrm{i}\partial_{t}+\mu\right]Q\left(t,t^{\prime}\right)-\left(\Delta\ast Q\right)\left(t,t^{\prime}\right)= & \delta_{\mathcal{C}}\left(t,t^{\prime}\right),\label{eq:Q}\\
\left[\mathrm{i}\partial_{t}+\mu-U\left(t\right)\right]R\left(t,t^{\prime}\right)-\left(\Delta\ast R\right)\left(t,t^{\prime}\right)= & \delta_{\mathcal{C}}\left(t,t^{\prime}\right)\label{eq:R}
\end{align}
where $\left(A\ast B\right)\left(t,t^{\prime}\right)=\int_{\mathcal{C}}d\bar{t}A\left(t,\bar{t}\right)B\left(\bar{t},t^{\prime}\right)$.
$\Delta\left(t,t^{\prime}\right)$ is the hybridization to a fictitious
bath. Note all the Green's function are defined on the Keldysh contour.
We use Eq.~(\ref{eq:R}) as an example to explain how to go to Floquet
space. Firstly, we write all functions explicitly on the two Keldysh
branches: 
\begin{equation}
\left(\Lambda\ast R\right)\left(t,t^{\prime}\right)-\left(\Delta\ast R\right)\left(t,t^{\prime}\right)=\delta_{\mathcal{C}}\left(t,t^{\prime}\right)
\end{equation}
where $\Lambda\left(t,\bar{t}\right)=\left(\mathrm{i}\partial_{t}+\mu-U\left(t\right)\right)\left(\begin{array}{cc}
\delta_{11}\left(t,\bar{t}\right)\\
 & \delta_{22}\left(t,\bar{t}\right)
\end{array}\right)$. Then we go to the real axis~\cite{rammer2007quantum}, 
\begin{equation}
\left(\tau_{3}\Lambda\ast\tau_{3}R\right)\left(t,t^{\prime}\right)-\left(\tau_{3}\Delta\ast\tau_{3}R\right)\left(t,t^{\prime}\right)=\tau_{3}\delta\left(t,t^{\prime}\right).
\end{equation}
With the Larkin-Ovchinikov transformation~\cite{Aoki2014}, we have
\begin{equation}
\left(\hat{\Lambda}\ast\hat{R}\right)\left(t,t^{\prime}\right)-\left(\hat{\Delta}\ast\hat{R}\right)\left(t,t^{\prime}\right)=\hat{\delta}\left(t,t^{\prime}\right)
\end{equation}
where we should note that $\hat{\Lambda}\left(t,\bar{t}\right)$ and
$\hat{\delta}\left(t,\bar{t}\right)$ are diagonal. The same procedure
can be applied to Eq.~(\ref{eq:Q}). We can transform these equations
to the frequency domain
\begin{align}
\hat{G}\left(\omega\right)= & w_{0}\hat{Q}\left(\omega\right)+w_{1}\hat{R}\left(\omega\right),\\
\hat{Q}_{mn}^{-1}\left(\omega\right)= & \left(\omega+\mu+n\Omega\right)\delta_{mn}-\Delta_{mn},\\
\hat{R}_{mn}^{-1}\left(\omega\right)= & \left(\omega+\mu+n\Omega\right)\delta_{mn}-U_{mn}-\Delta_{mn}.
\end{align}
These equations can be compactly written as 
\begin{equation}
\hat{G}\left(\omega\right)=w_0\hat{\mathcal{G}}_{0}\left(\omega\right)+w_1\left(\hat{\mathcal{G}}_{0}^{-1}\left(\omega\right)-\hat{U}\left(\omega\right)\right)^{-1}\label{eq:solver}
\end{equation}
where $\hat{U}\left(\omega\right)=\left(\begin{array}{cc}
\left(U_{mn}\right) & 0\\
0 & \left(U_{mn}\right)
\end{array}\right)$ with $\left(U_{mn}\right)$ a matrix consisting of elements $U_{mn}$.
Equation~(\ref{eq:solver}) also applies if $U\left(t\right)$ is
time-dependent. 

\subsection{\label{subsec:FGFM}Non-interacting retarded Floquet Green's function
for modulated interactions }

We write the Hamiltonian by explicitly including the chemical potential
$\mu$
\begin{equation}
H_{s}=-J\sum_{\left\langle ij\right\rangle }c_{i}^{\dagger}c_{j}-\mu\sum_{i}c_{i}^{\dagger}c_{i}+\left(U+\delta U\cos\left(\Omega t\right)\right)\sum_{i}c_{i}^{\dagger}c_{i}f_{i}^{\dagger}f_{i}.
\end{equation}
Imposing particle-hole symmetry at all time $t$, we find 
\begin{equation}
\mu=\frac{1}{2}\left(U+\delta U\cos\left(\Omega t\right)\right).
\end{equation}
One can transform the kinetic part to momentum space
\begin{equation}
H_{s}=-J\sum_{\bm{k}}\epsilon_{\bm{k}}\left(t\right)c_{\bm{k}}^{\dagger}c_{\bm{k}}+\left(U+\delta U\cos\left(\Omega t\right)\right)\sum_{i}c_{i}^{\dagger}c_{i}f_{i}^{\dagger}f_{i}
\end{equation}
where $\epsilon_{\bm{k}}\left(t\right)=-2J\sum_{\alpha=1}^{d}\cos\left(k_{\alpha}\right)-\frac{1}{2}\left(U+\delta U\cos\left(\Omega t\right)\right)$.
It has been shown~\cite{Aoki2008,Frank2013,Aoki2014} that for a
driven system the non-interacting retarded Green's function can be
expressed as 
\begin{equation}
G_{\bm{k}}^{R0}\left(\omega\right)=\Lambda_{\bm{k}}\cdot\left[Q_{\bm{k}}^{-1}\left(\omega\right)+\mathrm{i}\Gamma\right]^{-1}\cdot\Lambda_{\bm{k}}^{\dagger}
\end{equation}
where the term $\mathrm{i}\Gamma$ arises from the bath and 
\begin{align*}
\left(\Lambda_{\bm{k}}\right)_{mn} & =\int_{-\pi}^{\pi}\frac{dx}{2\pi}e^{\mathrm{i}\left(m-n\right)x}\\
 & \times\exp\left(-\frac{\mathrm{i}}{\Omega}\int_{0}^{x}dz\left[\epsilon_{\bm{k}}\left(\frac{z}{\Omega}\right)-\left(\epsilon_{\bm{k}}\right)_{0}\right]\right)
\end{align*}
 and $\left(Q_{\bm{k}}\right)_{mn}\left(\omega\right)=\frac{1}{\omega+n\Omega+\frac{U}{2}-\left(\epsilon_{\bm{k}}\right)_{0}+\mathrm{i}\eta}\delta_{mn}$.
One obtains
\begin{align}
G_{\bm{k}mn}^{R\left(0\right)}\left(\omega\right) & =\sum_{l}\mathcal{J}_{l-m}\left(\frac{\delta U}{2\Omega}\right)\mathcal{J}_{l-n}\left(\frac{\delta U}{2\Omega}\right)\\
 & \times\frac{1}{\omega+l\Omega+\frac{U}{2}-\epsilon_{\bm{k}}+\mathrm{i}\Gamma}.
\end{align}
We next perform the integral over $\epsilon_{\bm{k}}$ for a Gaussian
DOS~\cite{Georges1996}, which leads to 
\begin{align}
G_{mn}^{R\left(0\right)}\left(\omega\right) & =\sum_{l}\mathcal{J}_{l-m}\left(\frac{\delta U}{2\Omega}\right)\mathcal{J}_{l-n}\left(\frac{\delta U}{2\Omega}\right)\\
 & \times\left(-\mathrm{i}\sqrt{\pi}e^{-\xi_{l}^{2}}\mathrm{erfc}\left(-i\xi_{l}\right)\right)
\end{align}
where $\xi_{l}=\omega+l\Omega+\frac{U}{2}+\mathrm{i}\Gamma$ and $\mathrm{erfc}\left(z\right)\equiv\frac{2}{\sqrt{\pi}}\int_{z}^{\infty}e^{-t^{2}}dt$. 

\bibliographystyle{apsrev4-1}
%

\end{document}